\documentclass[12pt]{article}

\usepackage{xcolor}

\def\hybrid{\topmargin 0pt      \oddsidemargin 0pt
	\headheight 0pt \headsep 0pt
	\textheight 9in         
	\textwidth 6.25in       
	\marginparwidth .875in
	\parskip 5pt plus 1pt   \jot = 1.5ex}

\catcode`\@=11
\def\marginnote#1{}
\newcount\hour
\newcount\minute
\newtoks\amorpm
\hour=\time\divide\hour by60
\minute=\time{\multiply\hour by60 \global\advance\minute by-\hour}
\edef\standardtime{{\ifnum\hour<12 \global\amorpm={am}%
	\else\global\amorpm={pm}\advance\hour by-12 \fi
	\ifnum\hour=0 \hour=12 \fi
	\number\hour:\ifnum\minute<10 0\fi\number\minute\the\amorpm}}
\edef\militarytime{\number\hour:\ifnum\minute<10 0\fi\number\minute}

\def\draftlabel#1{{\@bsphack\if@filesw {\let\thepage\relax
   \xdef\@gtempa{\write\@auxout{\string
      \newlabel{#1}{{\@currentlabel}{\thepage}}}}}\@gtempa
   \if@nobreak \ifvmode\nobreak\fi\fi\fi\@esphack}
	\gdef\@eqnlabel{#1}}
\def\@eqnlabel{}
\def\@vacuum{}
\def\draftmarginnote#1{\marginpar{\raggedright\scriptsize\tt#1}}

\def\draft{\oddsidemargin -.5truein
	\def\@oddfoot{\sl preliminary draft \hfil
	\rm\thepage\hfil\sl\today\quad\militarytime}
	\let\@evenfoot\@oddfoot \overfullrule 3pt
	\let\label=\draftlabel
	\let\marginnote=\draftmarginnote
   \def\@eqnnum{(\theequation)\rlap{\kern\marginparsep\tt\@eqnlabel}%
\global\let\@eqnlabel\@vacuum}  }

\def\numberbysection{\@addtoreset{equation}{section}
	\def\theequation{\thesection.\arabic{equation}}}

\def\underline#1{\relax\ifmmode\@@underline#1\else
	$\@@underline{\hbox{#1}}$\relax\fi}

\def\titlepage{\@restonecolfalse\if@twocolumn\@restonecoltrue\onecolumn
     \else \newpage \fi \thispagestyle{empty}\c@page\z@
	\def\thefootnote{\fnsymbol{footnote}} }

\def\endtitlepage{\if@restonecol\twocolumn \else  \fi
	\def\thefootnote{\arabic{footnote}}
	\setcounter{footnote}{0}}  
\catcode`@=12
\relax
 
\def\demi{{1\over 2}}

\def\ee{\eea}
\def\be{\bea}

\def\beq{\begin{equation}}
\def\eeq{\end{equation}} 
\def\bea{\begin{eqnarray}}
 \def\eea{\end{eqnarray}}

\def\bar{\overline}
 \def\z{{\bar {z}}}
 \def\nn{\nonumber}

\def\t{{\theta}}

\def\P{\Psi}

 \def\demi{{1\over 2}}

\relax
\numberbysection
\hybrid
\begin{document}
  
 \def\bbr  {\textcolor{red}}    
 \def\bbr  {\textcolor{black}}

\title{{{\Large{\bf
An Angular  \bbr{Dependent }   Supersymmetric \\Quantum Mechanics  with a ${\bf Z}_2$-invariant Potential}}}}

\author{{
{Laurent Baulieu}}\thanks{{ E-mail: {baulieu@lpthe.jussieu.fr}}}
{\quad and\quad} {{Francesco
Toppan}}\thanks{{ E-mail: {toppan@cbpf.br}}}
\\
}
\date{}
\maketitle
\centerline{$^{\ast}$
{\it   LPTHE, Sorbonne	 Universit\'e and CNRS UMR 7589,}}
{\centerline{\it 
	4 place Jussieu,
 75005 Paris, France.}}
\centerline{$^{\dag}$
{\it CBPF, Rua Dr. Xavier Sigaud 150, Urca,}}{\centerline {\it\quad
cep 22290-180, Rio de Janeiro (RJ), Brazil.}
\begin{abstract}
 We generalize the conformally invariant  topological quantum mechanics   of a particle  propagating on a punctured plane by introducing a potential that breaks both the rotational and  the conformal invariance down to a   ${\bf Z}_2$ angular-dependent discrete symmetry.
\bbr { We derive a topological quantum mechanics whose   localization  gauge functions give interesting self-dual equations. The model contains an order parameter and exhibits {\textcolor{black}{a critical phenomenon with the existence of}} two ground states above a critical scale.  Unlike the ordinary $O(2)$-invariant Higgs potential,
an angular-dependence is found and saddle points, instead of local maxima, appear, posing subtle questions
about the existence of instantons.  The supersymmetric   quantum mechanical model is constructed in both  the path integral and the operatorial frameworks. }
\end{abstract}

\vskip1cm
\hfill{CBPF-NF-007/17~~~~~~~}
\thispagestyle{empty}

\newpage
\section {Introduction }

In ~\cite{BR} a solvable model for a superconformal  quantum mechanical system was introduced,
giving  an example of a quantum topological theory
  with   no  ground state and no mass gap.   In  ~\cite{BT} this model was extended  
to the case of   a possibly infinite chain. This toy model was motivated by the search  of an abstract definition of   mechanical systems possessing epicycle descriptions, {in an attempt to encode the robotic movements of rods presented in ~\cite{vit}}.  {\textcolor{black}{In this work topological models of rigid bars were presented.
They admit zero-energy modes which do not stress or compress the components. Kink solutions are shown to propagate and, in a certain regime, a supersymmetric effective description is shown to emerge. The topological chains of ~\cite{BT}, with their epicycle movements, {\color{black}  {are defined by   a quantum mechanical potential which  }} naturally encodes several of these features, such as the fixed length of the bars and the supersymmetry. A more refined description is, nevertheless, required to take into account the existence of solitonic kinks and the restriction of the angular movements in a given range less than $2\pi$. This is the main source of inspiration for proposing, in this paper, a topological quantum mechanics with a ${\bf Z}_2$ discrete symmetry potential. }}\par
More refined  models are expected to have a breaking of conformal symmetry. It  is tempting to  look for new classes of potentials  that  break other types of symmetries as well, like the rotational invariance, while keeping some of the properties of the topological models ~\cite{BR,BT}. Here we generalize   the topological invariant $\int_\gamma  d \theta$ that defines the topological quantum mechanics of  ~\cite{BR}  
 {\textcolor{black}{into another invariant}}, which has a further dependence on a point $A$ distinct from  the singular origin $O$ of the punctured plane; this allows us to introduce a length $a=|{\overrightarrow{OA}}|$ and a dimensionless order parameter. The model is defined by a potential which breaks both the conformal and the rotational symmetry, but keeps  a discrete ${\bf Z}_2$ symmetry. This discrete symmetry is reminiscent of the $T$-duality transformation in string theory, see ~\cite{Tdual}. {\textcolor{black}{In this loose analogy the duality transformation involves not only a transformation based on $r$, but also a $\theta$ angular dependence of ``$B$-field" type.}} \par The introduction of the length scale allows for  the existence of a discrete spectrum  and a  normalizable ground state.  {\textcolor{black}{The richness of the model appears in the presence of a critical phenomenon. Indeed, above a critical value for the order parameter,
two ground states are found;      {\color{black} {below that value, the ground state is unique. }}}} Two main properties single out this potential with respect to the standard Higgs potential. The first one is the presence of an angular-dependence; the second one is the fact that, above the critical scale, the potential does not possess any local maximum, but only (two) saddle points.
The possibility, following ~\cite{coleman},  of the existence of instantons is investigated.

\section {A reminder of conformal topological mechanics}

The conformally invariant  topological quantum mechanics of ~\cite{BR} is defined  by the topological gauge-fixing following the BRSTQFT, i.e. localization, scheme of ~\cite{localisation} for the topological invariant  $  \int_\gamma  d \theta$ on a punctured plane with real coordinates $q^i$ ($i=1,2,  x\equiv q^1,  y\equiv q^2$) or complex coordinates $q^1 +i q^2 =  \sqrt {|
\vec{q}| ^2} \exp i\t  $.\par By
 introducing   a coupling constant $g$,  the topological classical action  ~\cite{BR} reads
\be \label{action1}
2\pi N  g=  g \int_\gamma  d \theta = g \int_\gamma   dt \  \dot \theta  = g\int_\gamma   dt
  \epsilon_{ij} \frac {q^i\dot q^j}{2{|\vec q|}^2 },
\ee
{\textcolor{black}{where $t$ is a Euclidean}} time and the dot means the time derivative $\frac{d}{dt}$.\par
This action must be gauge-fixed in a topological BRST invariant way.
Using       the complex coordinate notation   $z(t) \equiv q^1(t) +i q^2(t) $ ($\z =z^\ast$), the topological gauge-fixing equation was obtained in ~\cite{BR} \bbr{as}:
\be\label{top0}
 \dot z - \frac {ig}{\z} &=&0.
\ee
The \bbr{BRSTQFT} gauge  invariant  procedure  provides a conformally invariant supersymmetric action \bbr { which, once untwisted,   gives a ${\cal N}=2$ supersymmetry}. \par The solutions of (\ref{top0})
   are periodic  instantons $z_N(t)$,  where  $N$ is an integer. We have
\be
z_N&=&  \frac{1}{\sqrt {    2\pi N g}} \exp  (2\pi i N t).
\ee
\bbr{
They  minimize (to the zero value) the non-negative bosonic  action 
\be \int   _\gamma dt (\dot {{|\vec q}|^2}+\frac{g^2}{{\vec q}^2}   +  g\dot \t).
\ee
}
The instantons are  pseudo-particles that make   $N$  cycles  per unit time at constant angular velocity    around the singular origin     on   circles with radius
$
|z_N |=\frac{1}{
\sqrt {2\pi N g }
}
$. The   area $ \Sigma_N $ of the surface   that  the  instanton  $z_N$ circles around    is independent on the  value of $N$:
\be
\Sigma_N &=& N \pi
( \frac{1}{
\sqrt {2\pi N g }} )^2     = \frac{g}{2} .
\ee
{\textcolor{black}{The instanton structure  follows from  the identity
\be
 \dot |{\vec q}|^2+\frac{g^2}{|{\vec q}|^2} &
=&
   \frac{1}{2} ( {\dot q_i} - g  \epsilon_{ij} \frac {q^j}{\sqrt{2}|\vec q| ^2 })^2
   +
   \frac{1}{2} ( {\dot q_i} +g  \epsilon_{ij} \frac {q^j}{\sqrt{2}|\vec q| ^2 })^2,
\ee
where the vector indices $i,j$ are raised/lowered by the Euclidean metric $\delta_{ij}$.}}
\par
The equation of motion, derived from the supersymmetric action,  of the   fermionic partner
$\Psi=\Psi^1+i\Psi^2$  of $z $, corresponds to   the  supersymmetric variation  $Q$ of Eq. (\ref{top0}). We have\be\label{0mode}
Q( \dot z - \frac {ig}{\z} )&=&
ig\dot\Psi^* +\frac {g}{{|{\vec q}|^2}  }   U_\theta \Psi  =0.
\ee
Here $ U_\theta   =    R_{\theta}    C  R_{-\theta} $,  where $R_{\theta}$ represents  an $SO(2)$ rotation ($\det R_\t=1$) with angle $\t$ and $C$  is a conjugation  matrix with $\det C=-1$; therefore $\det  U_\theta=-1$.\par
 The representations of the matrices acting on vectors are
 \be\label{RCdef}
 &&C=
 {\small \pmatrix{
0 &1 
\cr
1&
0  
} 
}
,\quad \quad R_{\theta}={\small{
 \pmatrix{
 \cos  \t&
-\sin\t 
\cr
\sin\t&
\cos\t  
}
}},\nonumber\\
   && U_\theta   =|{\vec q}|^2{{\delta ^2 \t}\over{\delta q_i \delta q_j}} 
  = |{\vec q}|^2{{\delta ^2 \log  \sqrt {|{\vec q}|^2}}\over{\delta q_i \delta q_j}} =
 R_{\theta}    C  R_{-\theta}  
={\small
\pmatrix{
-\sin 2\t&
\cos 2\t 
\cr
\cos 2\t&
\sin 2\t  
} 
}.
\ee
In complex notation one has $U_\theta z =i{\textcolor{black}{\exp({ -2i\t})}}z^*$.\par
The action of the rotation $R_{\pm \t}$  on  the   instanton  $z_N$ 
is {\textcolor{black}{a}}  multiplication by   $\exp (\pm 2\pi i Nt)$.  
Therefore, the solution of the  fermion equation~(\ref{0mode})  in the $z_N$ background is 
\be
\Psi (t)  &=&    \Psi_0\ \exp( \pm  i2\pi Nt ),
\ee
where  $\Psi_0$  is time independent and  
\be
  \Psi^*_0 \pm   C  \Psi _0 &=&0.
\ee
It follows that $\Psi _0 $ is just an eigenvector of the operator $C$  which runs attached to the instanton. One   has    
\be
\Psi _0
&=&
{\small{ \pmatrix{
1 
\cr
\pm1 
} }}.
\ee
The fermionic zero mode{\textcolor{black}{s}} can be thought as  tangent vectors cycling     at constant frequencies $2\pi Ng$  around the circles of radius $\sqrt {2\pi Ng}$. Their  existence implies  a non-trivial cohomology for the BRST symmetry of the theory. {\textcolor{black}{The paper}}
 ~\cite{BR} discusses in detail the model and the way it  allows to compute some (simple)
 topological invariants   of the punctured plane. 

\section{A new model and its topological gauge-fixing}
\def\o{\omega}
\def\t{\theta}

Let us introduce a point $A\neq O$ on the punctured plane with singular origin $O$.
We define  the orthonormal coordinates ($x= q^1,\ y= q^2 $)  such that 
$ {\textcolor{black}{\overrightarrow{OA}}} =(a,0)$. We     also use  the complex and polar coordinates $z=q^1+iq^2=r\exp (i\t)$ when needed.

Given any   contour $\gamma$ \bbr {in the punctured plane}, we add to  the closed and not-exact form  $ g d \theta$ an exact-term that depend{\textcolor{black}{s}} on the distance of the particle to the point $A$. The simplest possibility is offered by 
$g d \theta   + \frac{\o}{2} d(|\vec q -  \overrightarrow{OA}|^2)$. We therefore define the classical topological action $I^{cl}_\gamma(g,\o)$ of the new model as
\be\label{action2}
I^{cl}_\gamma (g,\o) &=&\int_\gamma  \left(   g d \theta   + \frac{\o}{2} d(|\vec q -  \overrightarrow{OA}|^2)  \right).
\ee
The  action $ I^{cl}_\gamma (g,\o) $ is 
 independent of any infinitesimal shift  of coordinates
\bea\label{gauge1}
\delta r=\epsilon_r(t) , &&
\delta \t=\epsilon_\t(t),
\eea
with appropriate boundary conditions. 

Following the general \bbr{BRSTQFT} scheme of ~\cite{localisation}, we are interested in computing observables that satisfy the 
Ward identities corresponding to the above gauge symmetry~(\ref{gauge1}). We  will come back on this later, \bbr {by showing the possibility  of } {\textcolor{black}{ }} fermionic zero modes. Comparing the actions (\ref{action1}) and (\ref{action2}) {\textcolor{black}{one understands that the topological gauge-fixing (\ref{top0}) must be improved into}} 
\be\label{top1}
 \dot z &=&\frac {ig}{\z} +\o (z-a).
\ee
The above equation reads, in polar coordinates, as
\bea\label{polareq}
\dot r =\o  (r-a \cos \t   ),  &&
r\dot\t  =  \frac{g}{r} +\o a  \sin \t
\eea
and, in  Cartesian coordinates, as
\bea
\dot x  =     \frac{gy}{x^2+y^2}      +   \o  (x-a  ) , &&
\dot y =   -\frac{gx}{x^2+y^2}      +   \o  y.
\eea

The parametrization expressed by $a,\o,g$ can be simplified through redefinitions, \bbr{as we will shortly see}. On the other hand, to have the possibility of discussing several interesting limits,  \bbr {it is better to keep track of the explicit dependence on these parameters}. \par
The supersymmetric  $Q$-exact action that localizes the topological gauge fixing (\ref{top1}) is {\textcolor{black}{easier  to  compute} in Cartesian coordinates. It transforms the topological term into a supersymmetric action. It gives, on the same time, a consistent (that is, \bbr{BRSTQFT}-invariant) information {\textcolor{black}{about}} the $g,\o,a$ dependence of the theory. We get
\bea\label{susyac} I^{cl}_\gamma (g,\o) &=&\int_\gamma  \Big( g d \theta   + \frac{\o}{2} d(|\vec q -  \overrightarrow{OA}|^2)\mapsto  \nn\cr
\mapsto I^{susy}_\gamma (g,\o)
&=& \int_\gamma  \Big( g d \theta   + \frac{\o}{2} d(|\vec q -  \overrightarrow{OA}|^2) 
+  Q \Big (   \bar \Psi_x (\dot x -     \frac{gy}{x^2+y^2}      -   \o  (x-a  ) +\demi H_x  )  + \nonumber\\&&\quad\quad\quad\quad\quad\quad  +
\bar \Psi_y (\dot y +    \frac{gx}{x^2+y^2}     -  \o y +\demi H_y  )   \Big)\Big).
\eea
The time-dependent fields $\Psi_x(t) ,\Psi_y(t),\bar\Psi_x(t),\bar\Psi_y(t)$ are fermion (that is, Grassmann coordinates), while the time-dependent fields $H_x(t), H_y(t)$ are auxiliary bosonic fields satisfying algebraic equations of motion. The nilpotent operator $Q$  
($Q^2=0$, which also satisfies $[Q,  d/dt]=0$) acts as follows on the fields
\bea\label{qaction}
~~~~Qx=-\Psi_x, &\quad& ~Qy = - \Psi_y,\nonumber\\
Q\bar\Psi_x =H_x, &\quad & Q\bar \Psi_y = H_y
\eea
(and vanishing otherwise).\par
The computation of the $Q$-transformation of the topological gauge function provides a Dirac-type 
operator acting on fermions which, in Cartesian coordinates, reads as
\be
{\cal D}_{ij}&= & \delta_{ij}(\frac{d}{dt}-\o)  -g\frac {\epsilon_{ik}}{|{\vec q}|^2  }  (\delta_{jk} -2\frac{q^j q^k}     {|{\vec q}|^2  }).
\ee
Therefore
\bea
{\cal D}_{ij}&=&   \delta_{ij}(\frac{d}{dt}-\o)   +\frac {g }{|{\vec q}|^2  }
{\small{\pmatrix{
- \sin 2\t&
 \cos 2\t 
\cr
 \cos 2\t&
 \sin 2\t    
} _{ij}}}=
 \delta_{ij}(\frac{d}{dt}-\o)
+\frac {g }{|{\vec q}|^2  } (R_\t C R_{-\t} )_{ij}, \nonumber\\
&&
\eea 
where $R_\t$ and $C$ have been introduced in Eq. (\ref{RCdef}).
\par
We postpone to Section {\bf 7} some discussion about the supersymmetry of the model given by the action
$ I^{susy}_\gamma (g,\o)$.

\section{The bosonic part of the action}

We are interested at first in the vacuum and therefore in the  bosonic part of the Euclidean action $I^{susy}_\gamma (g,\o)$. After expanding the $Q$-exact term in (\ref{susyac}), eliminating the auxiliary fields  $ H_x,H_y$  by their algebraic  equations of motion and discarding the terms involving fermions, one 
obtains the following {\textcolor{black}{Euclidean}} bosonic  action (\textcolor{black}{written, for convenience, in polar coordinates).} One has
\be\label{Ibosonic}
I_{bosonic}
= \int _\gamma  dt (  
\demi \dot r^2 + \demi r^2 \dot\t ^2 +V(r,\t)),
\ee
\bbr{where    the potential  $ V(r,\t)$ is non-negative, being the sum of squares:}
\be\label{pot}
  V(r,\t)&=& \demi \o ^2 (r-a \cos \t   )^2  +
\demi ( \frac{g}{r} +\o a     \sin \t)^2=\nn\\
 &=&
\demi\frac{g^2}{r^2} +\demi\o^2r^2
+ \frac{a g\o }{r  }  \sin\t
 -\o^2 ar\cos\t + \demi\o^2 a^2 .
\ee
\bbr {The potential  possesses the interesting ${\bf Z}_2$ discrete symmetry   $T$}} ($T^2={\bf I}$), given by
\be
\frac {g}{r} \leftrightarrow   {\omega}{r} , && \cos  \t  \leftrightarrow  -\sin \t  .
\ee
The symmetry can be expressed as
\be \label{sym}
  T& :&  \quad  {r}  \mapsto    r'=\frac {{r_0}^2}{  r}   ,  \quad\quad \t \mapsto \t'=-\t -  \frac {\pi}{2} ,
\ee
where $r_0$, given by
\be \label{rzero}
r_0 =\sqrt {\frac {g}{\o  } } ,
\ee
can be set as a unit of length. One can thus  define the dimensionless parameter
\be\label{alpha}
   \alpha \equiv \frac{a}{r_0}=    \sqrt {\frac {a^2\o}{g } } .
  \ee
The potential (\ref{pot}) can be expressed as the product of the dimensional factor $g\o$ times a dimensionless function: 
\be\label{pot2}
  V(r,\t)&=&g\o  \left (   \      
\demi\left(\frac{r_0 }{r}\right )^2+
\demi\left(\frac{r }{r_0}\right) ^2
+\alpha     \left (
 \frac{r_0 }{r  }  \sin\t
- \frac{r  } { r_0 } \cos\t \right)
+ \demi   \alpha  ^2 
\right).
\ee

{{{\color{black}{
The action (\ref{Ibosonic})  is rotationally invariant  at $\alpha= 0$. The rotational symmetry is broken for  $\alpha\neq 0$. On the other hand the discrete symmetry $T$  in (\ref{sym}) of  the potential is maintained.  We will show that the potential  exhibits a transition between a simple and a double well, with a critical value for the  dimensionless parameter $\alpha$.\par
 If we use $ {r_0 }$ as the   unit of length the only relevant parameter of the model is $\alpha= a$.  
In the limit $\alpha\rightarrow 0$ the model is mapped onto the Lagrangian with
Calogero potential supplemented by the De Alfaro Fubini Furlan oscillatorial term ~\cite{dff} (the potential being proportional to $r^2+1/ r^2$). In the absence of the oscillatorial term the model corresponds to the ~\cite{cal}  $SL(2)$ conformal mechanics. In the presence of the oscillatorial term the action is also $SL(2)$-invariant, but the
$SL(2)$ field transformations are trigonometric and carry the dependence on the dimensional parameter (see
~\cite{{{pap},{ht}}} for details).  The superconformal topological model ~\cite{BR} can be regarded as a topological twisted supersymmetric version, see ~\cite{bht}, of the ${\cal N}=2$ superconformal mechanics introduced in~
\cite{fura}. Its ${\cal N}=2$ superconformal symmetry closes an $sl(2|1)$ superalgebra, whose generators are the
$sl(2)$ subalgebra generators $H,D,K$, the supersymmetry operators $Q_1,Q_2$, its conformal superpartners
${\widetilde Q}_1, {\widetilde Q}_2$ and a $u(1)$ $R$-symmetry generator (see \cite{bht} for details).\par
In the presence of the oscillatorial term the spectrum is gapped and (at $\alpha=0$)  
 { {{\color{black}  its discrete energy levels are equally spaced.} } A geometrical interpretation for $\alpha$ is that it represents the distance, expressed in units of $r_0$, between the rest point of  the harmonic   oscillator and the singularity of the potential~$1/r^2$. 
\par
{\color{black} As a final remark we mention that for $\omega\neq 0$ we have one or two distinct classical vacua (depending on the value of $\alpha$) and normalized quantum mechanical ground states. In the 
 $\omega\rightarrow 0$ limit the quantum 
vacuum disappears because it it is no longer normalizable; in this limit the conformal invariance  holds true. }
{}
}    }}}

\section{The shape of the potential. Are there instantons?}

We follow the method explained by Coleman in ~\cite{coleman}. The potential (\ref{pot}) is a sum of squares. The instantons are  solutions of  the  Euclidean equations of motion giving a finite value to the action.   Since they must make it extremal, and since the action~(\ref{Ibosonic}) is non-negative,    they are solutions of  Eq.~(\ref{top1}). Avoiding the possibility   of bounces, they should connect   pairs  of  local maxima  $M^\pm$  of the Euclidean  potential $-V(x,\t)$, with 
  $  V(r^+,\t^+)  =V(r^-,\t^-)     $,     leaving  and reaching    $M^\pm$ at zero velocity. \par
In quantum mechanics with more than one degree of freedom one is often concerned with the stability of the Euclidean solutions, even when they are energetically possible for connecting two vacua. We should point out that in the present case the potential (\ref{pot2}), whose introduction is justified by geometrical considerations, leads to self-dual equations (\ref{polareq}) that cannot be solved analitically.
We also point out, depending on the order parameter (the coupling constant) of the model, that extrema of the potentials are local minima, local maxima and saddle points. Despite these difficulties, it is instructive to analyze the possibility for instantons and to draw conclusions about the classical solutions by analyzing the shape of the potential. \par
Therefore we  check at first under which condition two $M^\pm$ maxima are obtained.
The    Euclidean potential   $ -V(r,\t)$ is always negative or null, {\textcolor{black}{with 
$V\to \infty$}} when  $r\to\infty$ and $r\to 0$. {\textcolor{black}{It is obvious}} that $ -V(r,\t)$, for continuity reasons due to the behaviour of the potential  at  $r=\infty$ and $r=0$, has at least one  absolute maximum.
The two equations from   Eq.~(\ref{top1}) give, at
  $\dot r=\dot \t =0$, the condition for an instanton to start at zero velocity from a local maximum of the potential:
\bea\label{instanton}
 \o  (r-a \cos \t )  = 0,  &\quad&
   \frac{g}{r} +\o a     \sin\t =  0.
\eea
At $\o\neq 0$, $a\neq0$, these equations are equivalent to
\bea\label{eqq}
&&r=a \cos \t ,     \quad\quad \sin 2\t =   -\frac {2g}{\o a^2} , \quad\quad
\tan  \t = -\frac {g}{\o r^2}. 
\ee
It is convenient to express all results in terms of the two fundamental parameters, the unit length $r_0$
introduced in (\ref{rzero}) and the non-negative dimensionless constant $\alpha$ ($\alpha\geq 0$) introduced
in (\ref{alpha}). \par
By using this parametrization, the second equation in  (\ref{eqq}), since $|\sin 2\t|\neq 1 $, admits solution only in the range   
\bea
\alpha &\geq \sqrt{2}.
\eea
Furthermore, only two maxima $M^\pm$ of the potential $-V(r,\theta)$, with $V(M^\pm)=0$, can exist above the critical
value $\alpha_c=\sqrt{2}$ for $\alpha$.\par
We give the explicit expression of the two maxima $M^\pm$, both in polar ($r^\pm, \t^\pm$) and in Cartesian
($x^\pm, y^\pm$) coordinates. It is convenient to introduce the angle $\varphi$,
\bea
\varphi&=& \frac{1}{2}\arcsin(\frac{2}{\alpha^2}),\quad\quad 0<\varphi\leq \frac{\pi}{4}.
\eea
In terms of $\varphi$ we can write, in polar coordinates,
\bea\label{Mpmpolar}
r^+=\alpha r_0\cos\varphi,  &\quad& \theta^+ =-\varphi,\nonumber\\
r^-=\alpha r_0\sin\varphi,  &\quad& \theta^-=\varphi-\frac{\pi}{2}.
\eea
In Cartesian coordinates we have
\bea
x^+=\alpha r_0\cos^2\varphi,  &\quad& y^+ =-\frac{1}{2}\alpha r_0\sin(2\varphi)=-\frac{r_0}{\alpha},\nonumber\\
x^-=\alpha r_0\sin^2\varphi,  &\quad& y^-=-\frac{1}{2}\alpha r_0\sin(2\varphi)=-\frac{r_0}{\alpha}.
\eea
Due to the discrete $T$ symmetry of the potential, the following relations are found:
\bea
x^++x^- = \alpha r_0,\quad\quad &\quad&y^+=y^-,
\eea
as well as 
\bea
\sqrt{r^+r^- }=r_0, \quad &\quad&\quad \theta^++\theta^-= -\frac{\pi}{2}.
\eea
In order to further clarify the structure of the theory we compute the extremal points of the potential (\ref{pot2}), obtained by solving
the coupled system of equations
\bea
\frac{\partial V}{\partial\theta} =0 &\rightarrow&\frac{r}{r_0} \cos\theta +\frac{r}{r_0}\sin\theta =0,\nonumber\\
\frac{\partial V}{\partial r} =0 &\rightarrow&r^4-r_0^4-\alpha r_0r(r_0^2\sin\theta+r^2\cos\theta) =0.
\eea
At the extrema, we check the sign of the second-order derivatives to determine local minima, maxima and saddle points.
We present the complete set of solutions in the different ranges of the order parameter $\alpha$:\\
{\em i}) at $\alpha =0$ we recover the $O(2)$ rotational invariant theory. The case is analogous to the one investigated in ~\cite{BR}, with the important difference that the harmonic term allows for a discrete spectrum with a normalizable vacuum. The unbroken $O(2)$ symmetry implies a ${\bf S}^1$ circle of degenerate maxima. It is given by the points at distance $r_0$ from the origin and $-V(r_0,\theta)<0$; \\
{\em ii}) in the $0<\alpha<\sqrt{2}$ range two extremal points are found (we call them $M^{0a}$ and $M^{0b}$). Their polar coordinates (which do not depend on $\alpha$) are
\bea\label{M0ab}
M^{0a} \equiv (r_0,-\frac{\pi}{4}), &\quad & 
M^{0b} \equiv (r_0,\frac{3\pi}{4}).
\eea
By inspecting the second-order derivatives in $M^{0a}$, $M^{0b}$, one proves that $M^{0a}$ is the unique maximum of $-V(r,\theta)$, while $M^{0b}$ is a saddle point. We have 
\bea
-V(M^{0b})<-V(M^{0a})<0;
\eea\\
{\em iii}) at  $\alpha = \sqrt{2}$, the critical case, we have two extremal points as before. This time 
$-V(M^{0\alpha})=0$;
\\
{\em iv}) in the range $\alpha>\sqrt{2}$ we have four extremal points, the two maxima $M^{\pm}$ given by
Eq. (\ref{Mpmpolar}) and the two points $M^{0a}$, $M^{0b}$ given by Eq. (\ref{M0ab}). The two maxima are such that $-V(M^\pm)=0$, while in this range of $\alpha$ both $M^{0a}$, $M^{0b}$ are saddle points.\par
The value of the potential $V(r,\theta)$ from Eq. (\ref{pot2}), computed at $M^{0a}$, $M^{0b}$, reads
\bea
V(M^{0a,b}) &=& g\o (1+\frac{1}{2}\alpha^2\pm\alpha\sqrt{2}),
\eea
where the sign $-$ ($+$) corresponds to $M^{0a}$ ($M^{0b}$). \par
Energetically one has the possibility of a finite action trajectory linking $-V(M^\pm)$ and passing through $M^{0a}$
and/or $M^{0b}$. It is an open question, left for future investigation, whether such a trajectory indeed exists. 

\def\o{\omega}
\section{The fermionic part of the action}
\def\P{\Psi}
The fermionic part of the $ I^{susy}_\gamma (g,\o)$ action (\ref{susyac}) is 
\be
I_{fermion}&=&-\int dt
\pmatrix{
 \bar \P_x &
 \bar \P_y 
} Q{\small{\pmatrix{
\dot x -   \frac {gy}{x^2+y^2}-\o(x-a) 
\cr
\dot y +  \frac {gx}{x^2+y^2}-\o y 
}}} .
\ee
By taking into account Eq. (\ref{qaction}) for the operator $Q$ one has
\be
I_{fermion}&=&\int dt
\pmatrix{
 \bar \P_x &
 \bar \P_y 
}   {\small{\pmatrix{
\frac {d}{dt} 
    +  g \frac {2xy}{(x^2+y^2)^{2}}-\o  &
-g\frac {x^2-y^2}{(x^2+y^2)^{2}}
\cr
-g \frac {x^2-y^2}{(x^2+y^2)^{2}}&
\frac {d}{dt} -
     g \frac {2xy}{(x^2+y^2)^{2}}-\o
}}}{\small{\pmatrix{\P_x
\cr
\P_y
} }}.
\ee
We call this expression the ``classical fermionic action"; the fermions $\Psi_x,\Psi_y, {\bar \Psi}_x,{\bar\Psi}_y$ are time-dependent anticommuting Grassmann variables.\par
The fermionic zero modes are defined as the solutions of the following equation, 
 \bea
{\small{ \pmatrix{
\dot\P_x
\cr
\dot\P_y 
} }}
&=&
{1\over{r^2}}
{\small{\pmatrix{
-g\sin 2\t     +\o r^2&
g\cos 2\t 
\cr
g\cos 2\t&
g\sin 2\t   +\o r^2
} }}{\small{ \pmatrix{
 \P_x
\cr
 \P_y 
} }}=
\nn\\
&=&
{1\over{r^2}}{\small{\pmatrix{
 \cos  \t&
-\sin\t 
\cr
\sin\t&
\cos\t  
}}}{\small{
\pmatrix{
\o r^2 &
 g 
\cr
 g&
\o r^2 
} }}{\small{
\pmatrix{
 \cos  \t&
\sin\t 
\cr
-\sin\t&
\cos\t  
}}}{\small{\pmatrix{
 \P_x
\cr
 \P_y 
} }}.
\eea

{\textcolor{black}{By setting $\P=\P_x +i\P_y$ one can also write}}  
\be
\dot \P&=&i \frac {g} {r^2} \exp( i2 \t)  \   \P ^*  +\o \P. 
\ee
In the above equation the functions  $r(t) $ and $\t(t)$ solve the self-duality equations (\ref{polareq}). 
\par
By redefining  $\P \to {\widetilde\P}=\exp( -i\t)\P$, one gets the equation
\be
\dot {\widetilde\P} +i\dot\t{\widetilde \P} =i \frac {g} {r^2}   {\widetilde  \P} ^*  +\o {\widetilde \P} .
\ee
By expressing $\dot\t$ in terms of the second equation of (\ref{polareq}), one has
\def\o{\omega}
\be
\dot\t  &=& \frac{g}{r^2} +\frac{\o a}{r}\sin\t,
\ee
so that one gets the following zero mode equation
\be
\dot {\widetilde\P}  + ( i \frac{g}{r^2} +i \frac{\o a}{r}\sin\t -\o ){\widetilde\P} &=&i \frac {g} {r^2}  {\widetilde \P}^*  .
\ee
\bbr{This first order differential equation indicates that, most likely,  fermionic zero modes exist in the background of the $r(t), \t(t) $ solutions of the self-duality equations (\ref{polareq}). }
\section{The supersymmetric Hamiltonian}

The passage from the supersymmetric Lagrangian entering (\ref{susyac}) to the classical Hamiltonian formulation is done following the standard prescription, by performing a Legendre transformation for both bosonic and fermionic coordinates. One introduces the conjugate momenta $p_i$ to the coordinates $q_i$, while the fermionic Grassmann fields ${\bar \Psi}_i$ turn out to be conjugate to the fermionic fields $\Psi_i$. The Poisson brackets are ${\bf Z}_2$-graded. The non-vanishing Poisson brackets
are
\bea\label{pbbrackets}
\{p_i,q_j\} = \delta_{ij}, &\quad& \{{\bar \Psi}_i,\Psi_j\}=\delta_{ij}.
\eea
The classical Hamiltonian $H$ is ${\cal N}=2$ supersymmetric. It corresponds to the Morse function
$S=g\t +\frac{\o}{2} |\vec q -\overrightarrow{OA}|^2 $. The supersymmetric charges $Q, {\bar Q}$ are given
by
\bea\label{charge}
Q=\Psi^i (  p_i -S_i), &&
\bar Q=\bar \Psi^i (  p_i +S_i),
\eea
where
 \be
 S_i &=& \frac{\delta S}{\delta q_i}=~  g  \epsilon_{ij} \frac {q^j}{{|\vec q| }^2 } +\o(q_i-a_i), \quad\quad (a_1=a, ~a_2=0).
\ee
One has
\be\label{classham}
H&=&\demi \{ Q,\bar Q\}= ~\demi p_i^2 -\demi  S_i^2
-\bar \Psi^i   \    \frac {\partial S_i}{\partial q^j}  \   \Psi^j.
\ee
A quick computation shows that the classical fermionic part of the Hamiltonian is
 \bea
\bar \Psi^i      (\frac {\partial S_i}{\partial q^j}) \Psi^j
&=&\o
(R_{-\t} \bar \Psi)^i 
 \Big(
{\bf I}_2 + \frac {r_0^2}{r^2}   
C
\Big )_{ij}  
(R_{-\t}  \Psi)^j,
\eea
where ${\bf I}_2$ is the $2\times 2$ identity matrix, while $R_{-\theta}, C$ have been introduced in (\ref{RCdef}). At this level  $\Psi^i$  and ${\bar \Psi}^i$ are Grassmann variables and one can compute correlation functions within the functional integral approach.\par
The quantum Hamiltonian ${\bf H}$ is obtained by realizing the ${\bf Z}_2$-graded Poisson brackets (\ref{pbbrackets}) in terms of (anti)commutators. This implies, in particular, that the quantum fermionic operators $\Psi^i$, ${\bar \Psi}^i$ are given by constant matrices satisfying Clifford algebra relations; for this reason extra terms are present with respect to the (\ref{classham}) expression of the Hamiltonian.\par
By performing an analytic continuation and a Wick rotation of the time coordinate, we can go back and forth from the Euclidean time to the real time formulation of the theory. We present here the results for the real time formulation. This means that the quantum Hamiltonian ${\bf H}$ is hermitian and the quantum supercharges ${\bf Q}$, ${\bf{\overline Q}}$ are hermitian conjugates:
\bea
{\bf H} ={\bf H}^\dagger,&\quad& {\bf Q}^\dagger={\bf{\overline Q}}.
\eea
A convenient presentation of the quantum fermionic operators $\Psi^i, {\bar\Psi}^i$ is via the $4\times 4$ matrices $\xi^{\pm i}$, through the positions
\be 
\xi^{\pm i} &=&\Psi^i\pm  \bar \Psi^i , 
\ee 
where $\xi^{\pm i}$ are given by
 \be
\xi^{+1}= 
{\small{\pmatrix{
 0 & 0 &1&0\cr
  0 & 0 &0&-1
  \cr
  1 & 0 & 0 & 0
  \cr
  0 & -1 & 0 & 0
}}} ,&&
\xi^{+2}= 
{\small{\pmatrix{
 0 & 0 &0&1\cr
  0 & 0 &1&0
  \cr
  0&1 & 0 & 0
  \cr
    1&0  & 0 & 0
} }},\nn\\
\xi^{-1}= 
{\small{\pmatrix{
 0 & 0 &1&0\cr
  0 & 0 &0&1
  \cr
  -1&0 & 0 & 0
  \cr
    0&-1 & 0 & 0
} }},&&
\xi^{-2}= 
{\small{\pmatrix{
 0 & 0 &0&1\cr
  0 & 0 &-1&0
  \cr
  0 & 1 & 0 & 0
  \cr
  -1& 0 & 0 & 0
} }}
.
\ee
The matrices $\xi^{\pm i}$ generate the $Cl(2,2)$ Clifford algebra. Indeed, by setting
\be
 \Gamma^\mu\equiv (\xi^{+1}, \xi^{+2},\xi^{-1},\xi^{-2}),
 \ee
for $\mu=1,2,3,4$, we end up with the basic relations, for their anticommutators,
 \be\label{salgebra}
 \{\Gamma^\mu,\Gamma^\nu\} =2 \eta^{\mu\nu}{\bf I}_4,
 \ee
where $\eta^{\mu\nu}$ is the flat metric with diagonal {\textcolor{black}{entries}} $(1,1,-1,-1)$. \par

The hamiltonian  ${\bf H}$ and the  supercharges ${\bf Q}, {\bf {\bar Q}}$ can be expressed as $4\times 4$  differential  matrix operators.  In polar coordinates the hamiltonian is
\bea\label{quantumham}
{\bf H}&=& \left(- \frac{1}{2}(\partial_r^2+\frac{1}{r}\partial_r+\frac{1}{r^2}\partial_{\t}^2)+V_0(r,\t)\right)\cdot {\bf I}_4 + {\bf V}_1(r,\theta),
\eea
where $V_0(r,\t)$ coincides with the potential in Eq. (\ref{pot}), namely
\bea
V_0(r,\t)&=&
\demi\frac{g^2}{r^2} +\demi\o^2r^2
+ \frac{a g\o }{r  }  \sin\t -\o^2 ar\cos\t + \demi\o^2 a^2,
\eea
while
\bea\label{V1}
{\bf V}_1(r,\t)&=&{\small{\pmatrix{
\o& 0 &0&0\cr
  0 & -\o &0&0
  \cr
  0 & 0 & -\frac{g}{r^2}\sin(2\t) & -\frac{g}{r^2}\cos(2\t) 
  \cr
 0&0 & -\frac{g}{r^2}\cos(2\t)  & \frac{g}{r^2}\sin(2\t) 
}}}.
\eea
The supercharge ${\bf Q}$ is explicitly given by
\bea\label{Qmatrix}
{\bf Q}&=&{\small{\pmatrix{
0& 0 &e_{13}&e_{14}\cr
  0 & 0 &0&0
  \cr
0&e_{32}  & 0& 0
  \cr
 0&e_{42}& 0& 0
}}},
\eea
with
\bea
e_{13}=-e_{42}&=& -i(\cos\t \partial_r-\frac{1}{r}\sin\t\partial_\t)-i\o (r\cos\t-a)+\frac{ig\sin\t}{r},\nonumber\\
~e_{14} =~e_{32}~&=&i(\sin\t\partial_r+\frac{1}{r}\cos\t\partial_\t)  +i\o r\sin\t+\frac{ig\cos\t}{r} .
\eea
The supercharge ${\bf {\bar Q}}$ is the hermitian conjugate of the matrix given in Eq. (\ref{Qmatrix}).
\par
The operators ${\bf H}$, ${\bf Q}$, ${\bf\bar Q}$ satisfy the ${\cal N}=2$ superalgebra
\bea
\relax \{{\bf Q}, {\bf\bar Q}\} = 2{\bf H}, && [{\bf H}, {\bf Q}]=[\bf H, {\bf \bar Q}]=\{{\bf Q},{\bf Q}\}=\{{\bf\bar Q},{\bf \bar Q}\}=0.
\eea
The Fermion Parity Operator ${\bf N_F}$ is given by the $4\times 4$ diagonal matrix
\bea
{\bf N_F} &=& diag(1,1,-1,-1).
\eea
Bosons (fermions) are the eigenvectors of ${\bf N_F}$ with eigenvalues $+1$ ($-1$).\par
We note that the Hamiltonian (\ref{quantumham}) cannot be diagonalized. Indeed, a rotation, acting on fermions alone,
which diagonalizes the potential ${\bf{V}}_1(r,\t)$, is $\theta$-dependent, so that it produces non-diagonal terms when applied to the Laplacian. By realizing that the lower two-dimensional block in ${\bf{V}}_1(r,\t)$ can be expressed as $-\frac{g}{r^2} R_{-\t}CR_\t$, where $R_\t, C$ have been introduced in (\ref{RCdef}), a simpler expression for the quantum dynamics is given by the hamiltonian ${\widehat {\bf H}}$, unitarily equivalent to ${\bf H}$ via the transformation
\bea\label{widehat}
{\widehat {\bf H}}&=& {\small{\pmatrix{
  {\bf I}_2 & 0 \cr
  0& R_{-\t} 
}}}  {\bf H} {\small{\pmatrix{
  {\bf I}_2 & 0 \cr
  0& R_{\t} 
}}}.
\eea 
One has
\bea\label{quantumhat}
{\widehat{\bf H}}&=& \left(- \frac{1}{2}(\partial_r^2+\frac{1}{r}\partial_r+\frac{1}{r^2}\partial_{\t}^2)+V_0(r,\t)\right)\cdot {\bf I}_4 + {\widehat{\bf V}}_1(r,\theta),
\eea
where the $4\times 4$ matrix  ${\widehat{\bf V}}_1(r,\theta)$ is decomposed in $2\times 2$ blocks according to
\bea
{\widehat{\bf V}}_1(r,\theta)&=&  {\small{\pmatrix{
  \o\sigma_3 & 0 \cr
  0&-\frac{1}{2r^2}({\bf I}_2+2(\partial_\t+g)C )
}}}  .
\eea
$\sigma_3$ is the diagonal Pauli matrix, while $C=\sigma_1$ is the conjugation matrix introduced in (\ref{RCdef}). The non-diagonal terms correspond to a first-order differential operator.

\section{Conclusions}

We introduced a deformation of the conformally invariant topological quantum model ~\cite{BR}  of a particle moving on the punctured plane.
The modified potential depends on a length scale  $r_0$ and on a dimensionless parameter $\alpha\geq 0$.
It explicitly breaks the rotational and the conformal invariance, but it preseves a discrete ${\bf Z}_2$ simmetry given by  the idempotent map  $r/r_0    \to r_0/r  $ and $\t\to-\t-\pi /4$ .}} The role of $\alpha$ is that of an order parameter with a critical value $\alpha_c=\sqrt{2}$. Below that value (for $0<\alpha\leq\sqrt{2}$) the ground state is unique. Above that value ($\alpha>\sqrt{2}$) the ground state is doubly degenerate.\par
Two key features distinguish this potential with respect to the well-known Higgs potential 
$ \lambda( | \vec q |^2 -\frac { m}{\lambda}   )^2$ in the plane. The first one is the presence of an angular dependence (since the model is not rotationally invariant). The second relevant feature is that, above the critical value,
the extremal points are the two ground states and two saddle points (therefore, the theory does not possess local maxima). \par
We proved (with techniques similar as those applied in ~\cite{coleman},~\cite{kin}) that, in the Euclideanized version of the model, instantons as defined by Coleman are energetically possible. The actual existence of the instantons, due to the fact that the self-duality equations are not analitically solvable and the potential has a complicated structure for the presence of saddle points, is an open question which deserves further investigations.\par
The BRSTQFT (see ~\cite{localisation}) gauge-fixing method induces a ${\cal N}=2$ supersymmetric quantum mechanics that we constructed both in the path-integral (the fermionic time-dependent fields are Grassmann coordinates) and in the operatorial (the fermionic degrees of freedom are realized by Clifford matrices) 
frameworks.\par
In the operatorial (and real time) formulation the Hamiltonian is a $4\times 4$ hermitian differential operator. By construction the Hamiltonian, for any $\alpha\geq 0$,  has a well-defined, discrete and bounded from below, spectrum. The Hamiltonian is non-diagonalizable and this is another indication of the richness and the non-triviality of the model.\par
We leave for a future paper the computation of the spectrum (numerical, if not analytical) of this quantum
theory and of its further properties.\par
As mentioned in the Introduction the construction of this model  with such a critical phenomenon was found  by searching for a potential that  can  possibly     simulate  the constraints of  a mechanical device that  acts as a series of locks, {\textcolor{black}{allowing  the propagation  of signals from  one side to the other  of a chain, reproducing the behaviour  of some of the machines described in ~\cite{vit}.}} 
~
\\ {~}~
\par {\large{\bf Acknowledgments}}
{}~\par{}~
{\textcolor{black}{
{L. B. is grateful to CBPF for hospitality and support from CNPq (BEV grant).
F. T. is grateful to LPTHE for hospitality. He received support from CNPq under PQ Grant 306333/2013-9 and
from  Labex ILP (reference ANR-10-LABX-63), Idex SUPER, part of program
Investissements d'avenir ANR-11-IDEX-0004-02.}}


\begin{thebibliography}{10}
 
 
 \bibitem{BR}  L. Baulieu and E. Rabinovici, {\it 
On the calculability of observables in topological quantum mechanical models}, 
  Phys. Lett. {\bf B 316} (1993) 93-101; arXiv:hep-th/9307067.
  
   \bibitem{BT} L.  Baulieu and F.  Toppan, {\it  Chains of topological oscillators with instantons and calculable topological observables in topological quantum mechanics},
   Nucl. Phys. {\bf B 912} (2016) 88-102; arXiv:1610.00943[hep-th]. 
   
   
\bibitem{vit} \bbr{V. Vitelli, N. Upadhyaya and B. G. Chen, {\it Topological mechanisms as classical spinor fields},
arXiv:1407.2890[cond-mat.soft].}

\bibitem{Tdual} A. Giveon, M. Porrati and E. Rabinovici, {\it Target Space Duality in String Theory}, Phys. Rept.
{\bf 244} (1994), 77-205; arXiv:hep-th/9401139.
 
    \bibitem{coleman} S. Coleman, {\it The Uses of Instantons}. In Zichichi A. (eds),  The Whys of Subnuclear Physics. The Subnuclear Series vol. 15, Springer, Boston, MA (1979) pages 805-941. 
Lecture delivered at Conference: C77-07-23 (Erice Subnucl. 1977:0805).

\bibitem{localisation} L. Baulieu and  I. M. Singer, 
{\it Topological Yang-Mills symmetry}, {Nucl. Phys. Proc. Suppl. \bf
5 B}, (1988), 12; {\it The topological sigma model}, {Comm. Math. Phys.  \bf 125}
(1989), 227; {\it Conformally invariant gauge-fixed actions for 2-D topological gravity}, {Comm. Math. Phys.  \bf 135}
(1991), 253. 
 
\bibitem{dff} V. de Alfaro, S. Fubini and G. Furlan, {\it Conformal invariance in quantum mechanics}, Nuovo Cimento {\bf  A 34} (1976) 569-612.


\bibitem{cal} F. Calogero, {\it Solution of a Three-Body Problem in One Dimension}, J. Math. Phys. {\bf 10} (1969) 2191-2197.

\bibitem{pap} G. Papadopoulos, {\it New potentials for conformal mechanics}, Class. Quant. Grav. {\bf 30}
(2013) 075018; arXiv:1210.1719[hep-th].

\bibitem{ht} N. L. Holanda and F. Toppan, {\it Four types of (super)conformal mechanics: D-module reps and invariant actions}, J. Math. Phys. {\bf 55} (2014) 061703; arXiv:1402.7298[hep-th].

\bibitem{bht} L. Baulieu, N. L. Holanda and F. Toppan, {\it A world-line framework for 1D Topological Conformal $\sigma$-models}, J. Math. Phys. {\bf 56} (2015) 113507; arXiv:1507.04995[hep-th].
{\textcolor{black}{\bibitem{fura} S. Fubini and E. Rabinovici, {\it Superconformal quantum mechanics}, Nucl. Phys. {\bf 
B 245} (1984) 17-44.}}

\bibitem{kin} Y. Z. Chu and T. Vaschapati, {\it Fermions on one or fewer kinks}, Phys. Rev. {\bf D 77} (2008)
025006; arXiv:0709.3668[hep-th].



 \end{thebibliography}
\end{document}